\documentclass{article}

\usepackage[utf8]{inputenc}
\usepackage{graphicx}
\usepackage{color}
\usepackage{xcolor}
\usepackage{float}
\usepackage{amsmath}
\usepackage{authblk}

\begin{document}

\title{Damage to Relativistic Interstellar Spacecraft by ISM Impact Gas Accumulation}

\author[1]{Jon Drobny}
\affil[1]{Department of Nuclear, Plasma, and Radiological Engineering, University of Illinois at Urbana Champaign, Urbana, IL, 61801, USA}

\author[2]{Alexander N. Cohen}
\affil[2]{Department of Physics, University of California - Santa Barbara, Santa Barbara, CA, 93106, USA}

\author[1]{Davide Curreli}

\author[2]{Philip Lubin}

\author[3]{Maria G. Pelizzo}
\affil[3]{Consiglio Nazionale delle Ricerche - Istituto di Fotonica e Nanotecnologie (CNR-IFN), 35131 Padova, Italy}

\author[4]{Maxim Umansky}
\affil[4]{Lawrence Livermore National Laboratory, Livermore, CA, 94550, USA}

\maketitle

\section{Abstract}

\textit{‘This is the version of the article before peer review or editing, as submitted by an author to the Astrophysical Journal. IOP Publishing Ltd is not responsible for any errors or omissions in this version of the manuscript or any version derived from it. The Version of Record is available online at [https://doi.org/10.3847/1538-4357/abd4ec]’.}

As part of the NASA Starlight collaboration, we look at the implications of radiation effects from impacts with the interstellar medium (ISM) on a directed energy driven relativistic spacecraft. The spacecraft experiences a stream of MeV/nucleon impacts along the forward edge primarily from hydrogen and helium nuclei. The accumulation of implanted slowly diffusing gas atoms in solids drives damage through the meso-scale processes of bubble formation, blistering, and exfoliation. This results in macroscopic changes to material properties and, in the cases of blistering and exfoliation, material erosion via blister rupture and delamination. Relativistic hydrogen and helium at constant velocity will stop in the material at a similar depth, as predicted by Bethe-Bloch stopping and subsequent simulations of the implantation distribution, leading to a mixed hydrogen and helium system similar to that observed in fusion plasma-facing components (PFC's). However, the difference in location of near-surface gas atoms with respect to the direction of exposure means that previously developed empirical models of blistering cannot be used to predict bubble formation or blistering onset. In this work, we present a model of the local gas concentration threshold for material blistering from exposure to the ISM at relativistic speeds. Expected effects on the spacecraft and mitigation strategies are also discussed. The same considerations apply to the Breakthrough Starshot mission.

\section{\label{sec:intro}Introduction}

\subsection{\label{sec:motivation}Motivation}

        If exploration beyond the outer limits of the solar system is to be conducted within reasonable fractions of a human lifetime, a radical shift in spacecraft propulsion technology is necessary. The radical shift proposed by Lubin \cite{Lubin2016AFlight} and the NASA Starlight project consists of the directed energy propulsion of ultra-light spacecraft and a low absorption reflective sail structure. The baseline system propels a low mass spacecraft via photon momentum exchange using a 1-10 km aperture, 100 GW class laser phased array to relativistic speeds. Such a system can be used with a wide variety of missions including  the acceleration of a 1 g spacecraft to $\sim$c/4 or a 10 kg spacecraft to $\sim$c/40  or greater. This would enable the first robotic interstellar journey to the nearest star system, $\alpha$ Centauri, with a flight time of less than 20 years. For this work, we use the term relativistic spacecraft to refer to this design and mission velocities greater than 0.1c. A spacecraft travelling at these speeds through the ISM would experience the ISM in its rest frame as a nearly mono-energetic beam of particles, consisting primarily of protons, electrons, and alpha particles. Thus, in designing a spacecraft capable of surviving interstellar journeys, it is crucial to characterize, understand, and mitigate the damage that is caused by incident particles to ensure spacecraft survival through the ISM.

\subsection{\label{sec:ISM}ISM Composition and Homogeneity}

        The ISM is composed of roughly $99\%$ gaseous matter and $1\%$ granular dust by mass. Of the gaseous matter, approximately $70\%$ is hydrogen and $28\%$ is helium by mass, while the remaining $2\%$ is made up of heavier elements including carbon, oxygen, and iron \cite{Klessen2016PhysicalMedium} \cite{Draine2011PhysicsMedium}. Interstellar dust is composed of still heavier constituents including hydrocarbons, silicates, and ices in the form of crystalline and amorphous grains of size less than $\sim$1 $\mu$m \cite{Draine2011PhysicsMedium}. Hydrogen in the ISM is observed in both neutral and ionized states, where the neutral state constitutes atomic and molecular hydrogen. The gaseous component of the ISM is commonly divided into five dominant phases which, since the phases are generally not in thermal equilibrium, drive large scale dynamics that produce structures like interstellar gas clouds of varying temperatures and often promote smaller scale gravitational phenomena such as star formation.

        On large scales ($\sim$10 to 100 ly), the ISM is inhomogeneous and can vary greatly in density. While the majority of the ISM by volume is composed of ionized hydrogen with a density of approximately 1 cm$^{-3}$, most of the neutral hydrogen is collected in large, cool clouds with densities between 0.1 and 50 cm$^{-3}$ for neutral atomic hydrogen clouds and often greater than 1000 cm$^{-3}$ for clouds cool enough to allow molecular hydrogen to exist in their cores \cite{Klessen2016PhysicalMedium} \cite{Draine2011PhysicsMedium}. These neutral gaseous phases are commonly referred to as cold and warm neutral media (CNM and WNM, respectively) depending on their temperatures, while the ionized phases are divided into the warm and hot ionized media (WIM and HIM, respectively).

        For relativistic spacecraft for which $\alpha$ Cen or any other local star is the target, the ISM can be approximated by measurements of the local interstellar cloud (LIC) due to the relatively short distance from the Sun to $\alpha$ Centauri. Common measurements of LIC densities, such as those performed by Gloeckler and Geiss, report combined H and He densities as low as $\sim$0.26 cm$^{-3}$ \cite{Gloeckler2004CompositionIons}. Therefore, for our analyses we conservatively approximate the local ISM as a homogeneous gas of atomic hydrogen with a density of 1 cm$^{-3}$ and a temperature of approximately 8,000 K. Thus, to a relevant first approximation, the local ISM falls within the WNM phase. However, distinct structures produced by the interaction of the ISM and stellar winds complicate the environment since they can create phases of matter that deviate greatly from the above approximation.

\subsection{\label{sec:stellarwinds}Stellar Winds}

        In addition to the various phases of the ISM that the spacecraft may experience, it may also be subject to the potentially damaging effects of stellar winds and the structures they produce as it approaches stellar targets. Specifically, in the simple stellar wind model consisting of a radial and isotropic outflow of material from the star, one would expect that, in the region where the stellar wind interacts strongly with the local ISM, there would exist a potentially different phase of material created by the interacting materials. In a NASA study by Linsky and Wood \cite{Linsky1996TheHeliopause}, it is described how the $\alpha$ Centauri system, though consisting of a total of three stars, likely produces a solar wind and astrosphere similar to the Sun's due to the proximity of the stars in the system, the main sequence nature of the two dominant stars, and the similarity in relative velocity to the local ISM of the Sun and the $\alpha$ Centauri system. It was found, and later verified by a number of groups \cite{Zank2013HELIOSPHERICWALL} \cite{Wood2004TheAstrospheres}, that in the region surrounding the $\alpha$ Centauri astropause, and thus likely also around the heliopause of the Sun and the astropause of most stars in the local cloud, there exists a heated region of ISM material compressed by the outflowing stellar wind to a temperature $\sim$30,000 K and density $\sim$0.3 cm$^{-3}$, commonly referred to as the hydrogen wall. It is interesting to note that at 0.25c, the spacecraft will likely pass through the hydrogen wall very quickly. Thus, we may be able to test this regime experimentally since the exposure time at hydrogen wall densities and temperatures may be small.

\subsection{\label{sec:spacecraftdesign}Relativistic Spacecraft Design}

\begin{figure}
            \centering \includegraphics[width=0.475\textwidth]{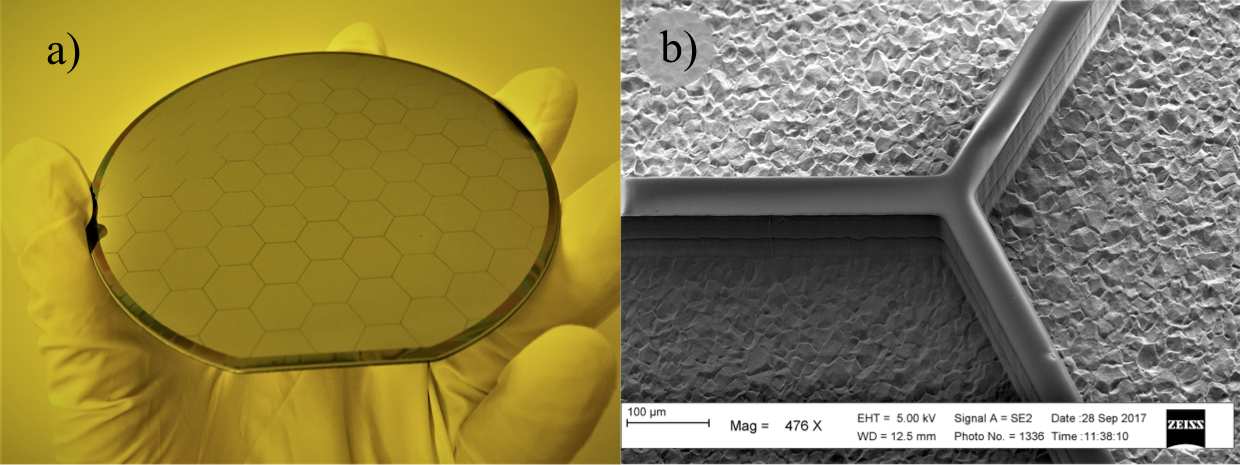}
            \caption{ \protect\centering a) Prototype 100 mm diameter silicon wafer scale spacecraft with a plasma etched honeycomb pattern to reduce the mass to 1 g. b) Detail of an etched honeycomb vertex. Fabricated at CNSI/UC Santa Barbara.}       \label{fig:honeycomb}
        \end{figure}

        The baseline design for small relativistic wafer scale spacecraft consists of a single thin semiconductor or hybrid disk of radius between 25 and 100 mm depending on the mission with a nominal diameter of 100 mm  \cite{Lubin2016AFlight}. The spacecraft flies edge on to reduce the number of ISM particle and dust impacts. The wafer houses the entire spacecraft bus, sensor suite, power system, and communication system on its surface, and would be manufactured from silicon or titanium using plasma etching technologies. It is to be extremely thin, $\sim$100 $\mu$m, with a thin-walled honeycomb structure on the other side to reduce mass and enhance its structural integrity, as shown in Figure \ref{fig:honeycomb}. The spacecraft wafer is then embedded within or attached to a light-sail structure made of an ultra-thin ($\sim$0.1 to 1 $\mu$m) dielectric, reflective and low absorption material that reflects the incident laser light, thus generating thrust. In other variants, the spacecraft could be a long thin ``needle-like'' structure with small forward-edge cross section. The lowest mass missions have a  wafer and reflector mass of order 1 g, though the same launch system is capable of larger missions with higher mass spacecraft  being slower with speed scaling as m$^{-1/4}$  \cite{Lubin2016AFlight}.

        In some proposed mission scenarios, the spacecraft design calls for a mechanism by which the reflector can be ejected from the wafer after the spacecraft has been accelerated to its cruise speed. In others, the reflector remains attached to the spacecraft for the duration of its journey or is removed passively by vaporization due to collisions with interstellar gas and dust.

        Due to the extremely large distances to interstellar mission targets, the spacecraft communicates with Earth using a dedicated laser communications system to send data back to Earth. The spacecraft quickly enters a regime where the command delay time becomes exceedingly long, requiring largely autonomous operation with infrequent uplink commands. The spacecraft is powered by the heat given off by a small pellet of radioactive material such as plutonium-238, though the ISM impacts could conceivably be used as a source of power. In most mission scenarios, the spacecraft enters a low power hibernation mode during transit to its destination. The details of this are discussed in our other papers \cite{Lubin2016AFlight}.

\section{\label{sec:cumulativeeffects}Cumulative Effects of ISM Impacts}

\subsection{\label{sec:ionmatter}Instantaneous Ion-Matter Interactions}

\begin{figure}
            \centering
            \includegraphics[width=0.45\textwidth]{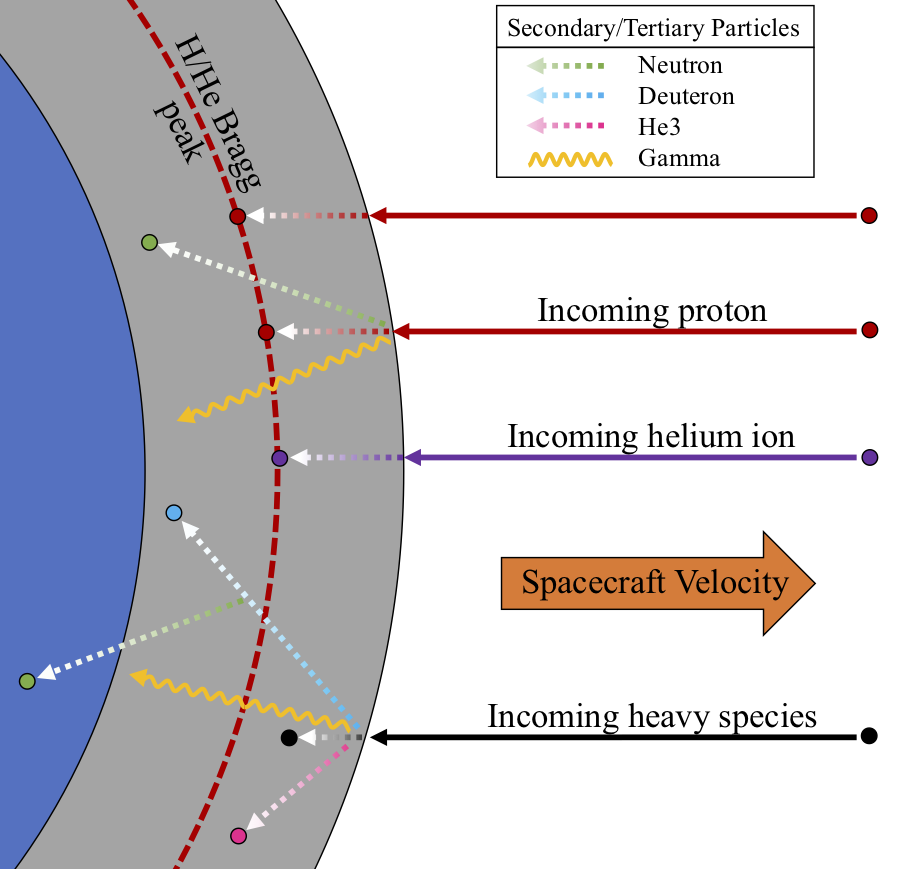}
            \caption{ \protect\centering Schematic of the various particle interactions in the plane of the spacecraft upon impact of ISM protons and heavier species on its leading edge. With exaggerated scales, the shield is shown in grey and the spacecraft wafer in blue. As shown, while the protons are stopped by the shield at the H/He Bragg peak, secondary and tertiary particles as well as heavier species have the capability of impinging far deeper into the spacecraft.}
            \label{fig:particles}
        \end{figure}

\begin{figure*}
            \centering
            \begin{tabular}{c c}
                \includegraphics[width=0.5\textwidth]{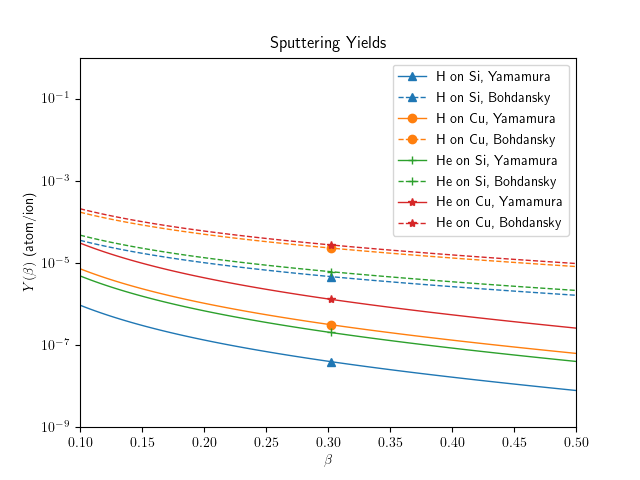} &
                \includegraphics[width=0.5\textwidth]{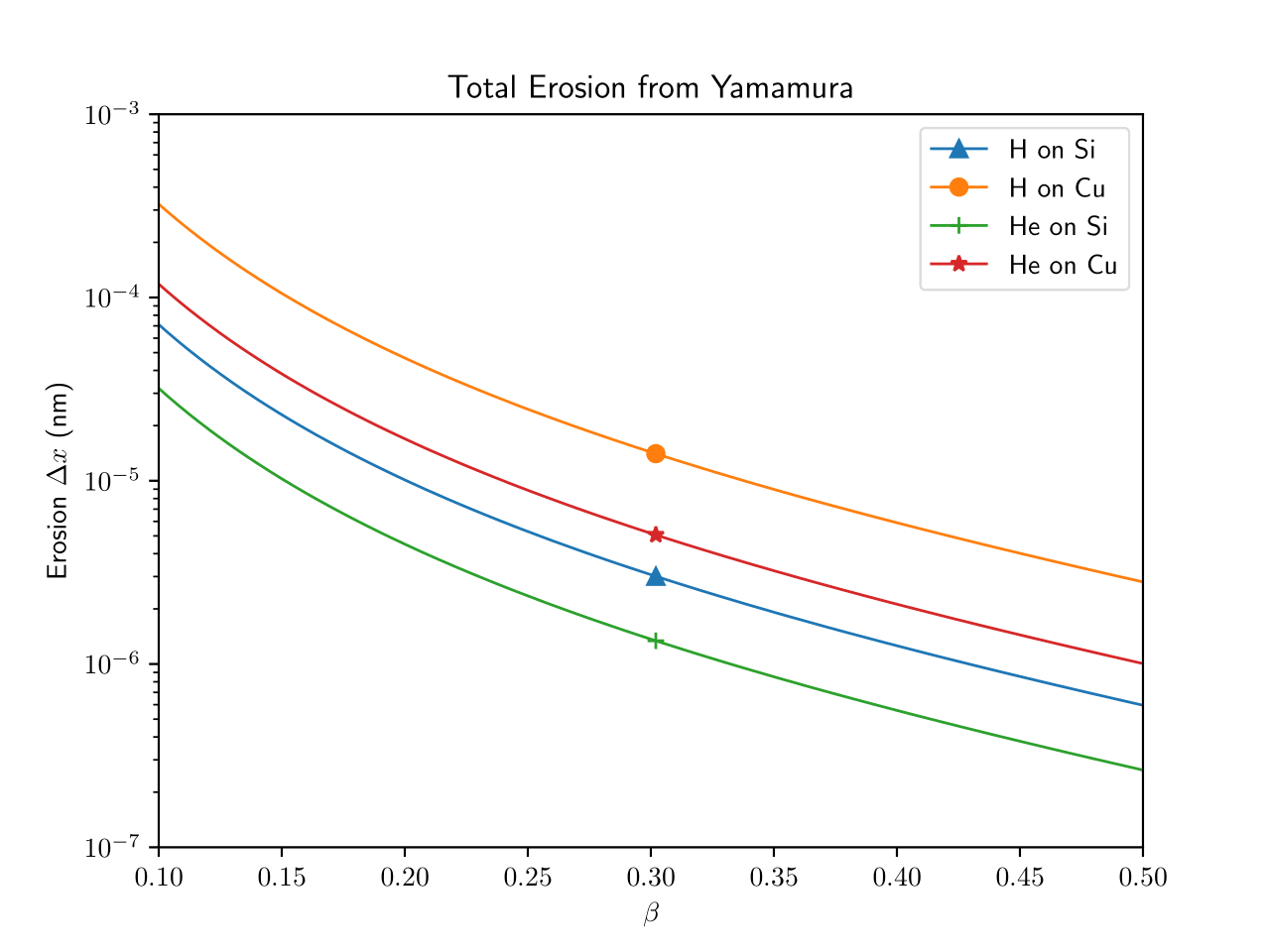} \\
                (a) & (b) \\
            \end{tabular}
            \caption{ \protect\centering (a) Sputtering yields from the Yamamura and Bohdansky formulas for hydrogen and helium incident on copper and silicon. (b) Total erosion from the Yamamura formula for sputtering yield, expected ISM fluence, atomic mass of the shield material, and mass density of the shield material, as shown in Equation \ref{eq:total_erosion}.}
            \label{fig:sputtering_yield}
        \end{figure*}

        ISM particle impacts will have a significant effect on relativistic spacecraft. Figure \ref{fig:particles} shows a summary of particle-spacecraft interactions, including hydrogen/helium implantation, heavy species impacts, and secondary particle generation. Since ISM particles incident upon a relativistic spacecraft will immediately lose their electrons on impact, we can treat the ISM in the frame of a relativistic spacecraft as a fully ionized, constant-velocity beam, composed primarily of hydrogen and helium. Ion-matter interactions span many orders of magnitude of energy-, time-, and length-scales and include the processes of sputtering, displacement of target atoms and damage, thermal spikes, and implantation. Contemporary understanding of these phenomena comes from a combination of theoretical, empirical, and computer models. Careful application of these models is necessary to understand the cumulative effects of ion bombardment. For example, a straightforward calculation of the analytic Bohdansky \cite{Bohdansky1984AIncidence} or semi-empirical Yamamura \cite{Matsunami1984EnergySolids} formulas for the sputtering yield of relativistic ISM impacts will predict a negligible amount of material erosion during an interstellar journey, as shown in Figure \ref{fig:sputtering_yield} (a).

        Total one-dimensional erosion depths can be calculated based on the fluence and target material parameters with the expression
        \begin{equation} \label{eq:total_erosion}
            \Delta x = \frac{\Phi_1 Y_{12}(E_1) M_2}{\rho_2},
        \end{equation}
        where $\Delta x$ is the depth of material eroded, $\Phi_1$ is the fluence of species 1, $Y_{12}(E_1)$ is the sputtering yield for species 1 incident on species 2, $M_2$ is the atomic mass of species 2, and $\rho_2$ is the mass density of species 2. Using this approximation, erosion of silicon and copper by hydrogen and helium for a journey to $\alpha$ Cen at speeds between 0.1 and 0.5c is less than a single atomic layer, as shown in Figure \ref{fig:sputtering_yield} (b). Erosion by physical sputtering is so low for two reasons: first, energy transfer during collisions between energetic light ions and heavy atoms is inefficient; second, most of the ion-atom energy transfer occurs deep below the surface, where the nuclear stopping cross section becomes significant compared to the electronic stopping cross section. For atoms to be sputtered, momentum must be transmitted from this location to a free surface. A previous work \cite{Hoang2017TheMedium} has considered the damage caused by the cumulative effect of instantaneous ion-matter interactions, including sputtering and track-formation. However, damage from the accumulation of gas atoms is an unexplored, potentially threatening phenomenon for relativistic spacecraft. Gas atom accumulation negatively affects material properties, causes material swelling, induces surface morphology changes through blistering, and can drive erosion through exfoliation at rates exceeding physical sputtering.

\subsection{\label{sec:bubbleformation}Bubble Formation and Blistering}

\begin{figure}
            \centering
            \includegraphics[width=0.45\textwidth]{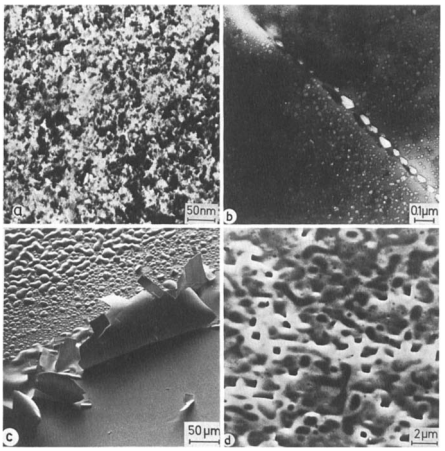}
            \caption{ \protect\centering Diagram showing damage to materials in the intermediate fluence range, including material defects, blistering, and exfoliation, shown here for illustrative purposes. Reprinted by permission from Springer-Verlag:  \cite{Scherzer1983DevelopmentImplantation}, 1983.
            }
            \label{fig:damage}
        \end{figure}

\begin{figure*}
            \centering
            \includegraphics[width=0.8\textwidth]{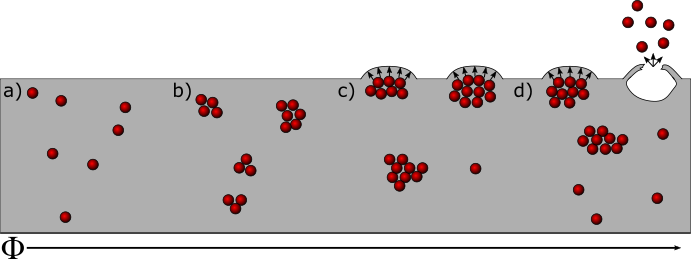}
            \caption{ \protect\centering Illustration of the process of bubble formation and blistering, neglecting gas atom interactions and material defects. a) At very low fluence ($<$ 10$^{18}$ m$^{-2}$), individual gas atoms are implanted in the material. They may diffuse freely and aggregate, or be trapped by defects. b) At low fluence ($\sim$10$^{18}$ m$^{-2}$), nano-scale bubbles will form in the material, but no surface effects will be visible. c) At intermediate fluence (10$^{18}$-10$^{21}$ m$^{-2}$) the local gas concentration will put stress on the material, and if at the surface, will begin to make morphological changes. d) At high fluence ($>10^{21}$ m$^{-2}$) surface blisters will begin to burst, leading to exfoliation, flaking, and ultimately, material erosion.}
            \label{fig:bubble_formation}
        \end{figure*}

        Bubble formation, blistering, and exfoliation are deleterious phenomena caused by the accumulation of relatively insoluble gas atoms in solids. Figure \ref{fig:bubble_formation} shows an illustration of the bubble formation and blistering process. Insoluble gas atoms implanted in a solid, such as helium, can aggregate and form overpressurized bubbles \cite{Donnelly1985TheReview}. If the local gas concentration exceeds a critical concentration, near-surface bubbles can deform the surface and burst, in processes called blistering and exfoliation. When blistering does not occur, such as in the case of a subcritical gas concentration, the implanted gas atoms negatively affect material properties and can lead to swelling and crack nucleation \cite{Condon1993HydrogenMetals}. Figure \ref{fig:damage} shows examples of surface damage caused by implanted gas atoms, including blistering and exfoliation.

         \begin{figure}
            \centering
            \begin{tabular}{c}
                \includegraphics[width=0.45\textwidth]{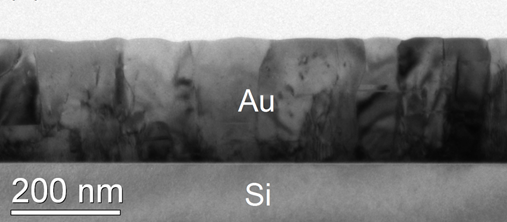} \\
                (a) \\
                \includegraphics[width=0.45\textwidth]{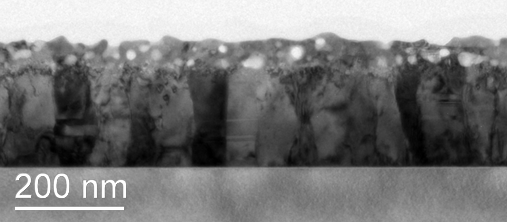} \\
                (b) \\
            \end{tabular}
            \caption{ \protect\centering (a) Gold sample prior irradiation.  (b) Gold sample after irradiation with 4 keV He ions with a total fluence of $4\times10^{17}$ cm$^{-2}$. Blistering is observed close to the surface at ions penetration depth. Reprinted with permission from  \cite{Pelizzo2018MorphologicalIons}. Copyright (2018) American Chemical Society.}
            \label{fig:bubbles}
        \end{figure}

        Blistering is a phenomenon typically encountered when materials are exposed to high-flux, high-fluence, intermediate energy (keV range), insoluble gas ions, such as in ion beam experiments and nuclear fusion \cite{Balden2013BlisteringPlasma} \cite{Behrisch2010Plasma-facingDevices} \cite{Bauer1978SurfaceInteractions}. Blistering by hydrogen and helium specifically has been extensively studied in these contexts \cite{Das1975RadiationAlloys} \cite{Kaletta1980LightMetals} \cite{St-Jacques1978HeliumMetals} \cite{Evans1976AIons} \cite{Erents1973BlisteringBombardment} \cite{Guseva1981RadiationBlistering}.
        Blistering induced by low energy protons has also been observed in extreme ultraviolet multilayer optics used in photo-lithographic systems \cite{vandenBos2017BlisterIons} and space instrumentation \cite{Pelizzo2011StabilityBombardment}, and in metal films tested for space environmental conditions \cite{Sznajder2018HydrogenConditions}. In Figure \ref{fig:bubbles} an example of blistering occurring in a thin gold layer is shown. The sample has been irradiated with 4 keV low energy helium ions with a total fluence of 4$\times$10$^{21}$ m$^{-2}$; at lower fluences, no clear evidence of bubbles was found. At higher energies (MeV range), blistering has been reported to occur at fluences an order of magnitude below the critical dose found from analogous experiments at intermediate energies (keV range). \cite{GavishSegev2017BlisterProtons}. In these situations, blistering occurs on the surface exposed to the ion flux, because the ions implant close to the surface.

        In the frame of a relativistic spacecraft, the ISM appears as a low-flux, high-fluence, MeV-energy beam that spans the entire front-facing cross section of the spacecraft. During relativistic travel, the implantation depth distributions of ISM gas atoms will be strongly peaked many atomic layers below the surface. Additionally, due to the scaling of Bethe-Bloch stopping, at a constant relativistic velocity, hydrogen and helium will stop at approximately the same depth. From the prefactor of the Bethe formula, $S_{0}$, shown in Equation \ref{eq:bethe} with $Z_{1}$ being the atomic number of the incident ion, $Z_{2}$ being the atomic number of the target material, $m_{e}$ being the electron mass, and $v$ being the incident velocity, helium will experience a stopping power approximately 4 times larger than hydrogen. However, since helium is approximately 4 times more massive than hydrogen, helium at constant velocity will have approximately 4 times more kinetic energy, resulting in similar ranges in a given material. This will result in a thin layer of material spanning the entire exposed portion of the spacecraft, at the depth of the implantation distribution, where the local concentration of mixed gas atoms may form bubbles and may exceed the critical concentration necessary to form blisters. The uniformity of the implantation across the entire front-facing cross-section suggests near-surface bubbles will form almost immediately on surfaces perpendicular to the ion flux.
        \begin{equation}
            \label{eq:bethe}
            S_{0} = \frac{4 \pi e^{4} Z_{1}^{2} Z_{2}}{m_{e} v^{2}}
        \end{equation}

\begin{figure*}
            \centering
            \begin{tabular}{c c}
                \hspace{-5mm}
                \includegraphics[width=0.4\textwidth]{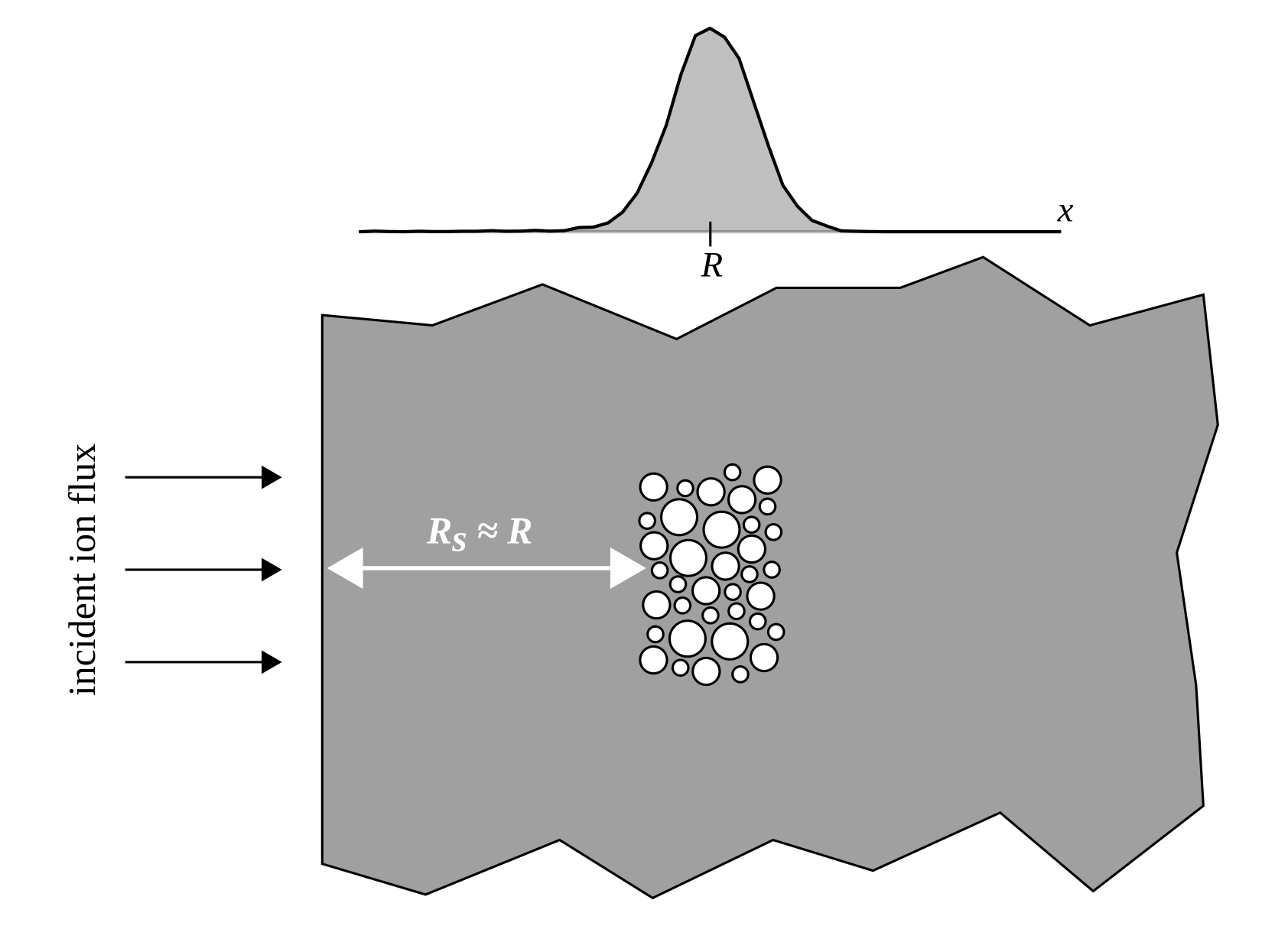} &
                \hspace{5mm}\includegraphics[width=.5\textwidth]{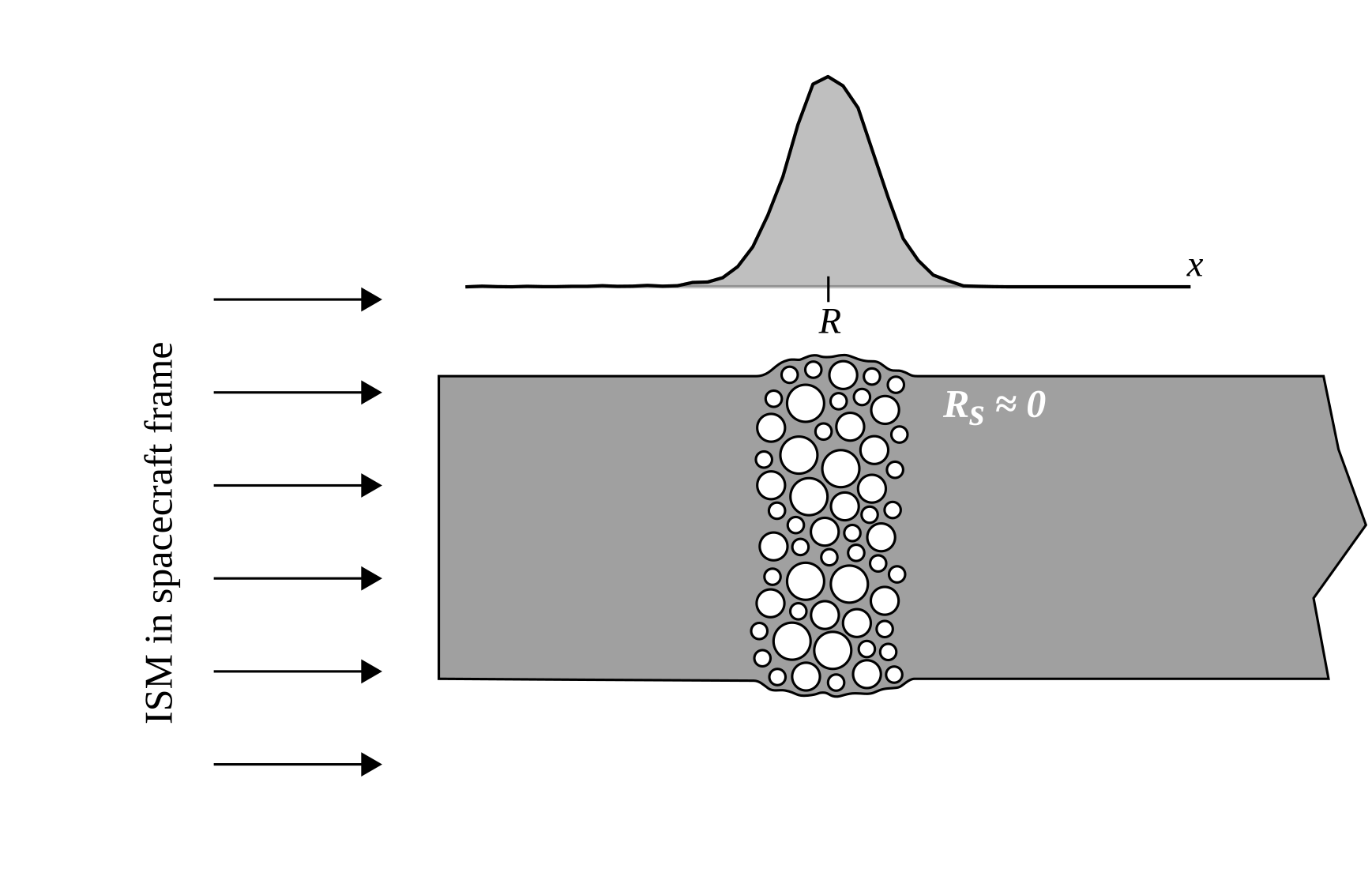} \\
                \hspace{10mm}(a) & \hspace{20mm}(b) \\
            \end{tabular}
            \caption{ \protect\centering (a) Bubble formation due to high energy ion beam exposure with beam width significantly narrower than target size. $R_s$, the distance from bubbles to the nearest surface, is close to the range of the ion implantation distribution, approximated above by a normal distribution. (b) Interstellar spacecraft configuration traveling edge on. $R_s$ is effectively zero because ISM exposure is spacecraft-spanning. Any swelling, blistering, or exfoliation will likely occur first on surfaces with normals perpendicular to the incident flux.}
            \label{fig:difference_versus_beam_experiments}
        \end{figure*}

        In Figure \ref{fig:difference_versus_beam_experiments} (a), an illustration of bubble formation during a high energy ion beam experiment is shown; the distance to the nearest surface from the location of bubble formation is approximately the range of the implantation distribution. In order to cause blistering or modify surface morphology, bubbles must travel to this surface to have an effect. Figure \ref{fig:difference_versus_beam_experiments} (b) shows bubble formation caused by ISM gas accumulation. In this case, the smallest distance from bubble formation locations to a surface is effectively zero; bubbles do not need to travel any distance to begin having a deleterious effect on the nearest surface. For this reason, swelling, blistering, and exfoliation may happen significantly earlier for the case of wide exposure. Previous work on blistering has produced an empirical formula \cite{GavishSegev2017BlisterProtons} for predicting the dependence of critical dose on the incident ion energy; since this formula assumes blister formation at the surface directly exposed to ion flux, it is not applicable here. Instead, a more fundamental model of blistering onset must be used.

        A simple theoretical model of blistering onset was described by Martynenko, and a reduced version is reproduced here \cite{Martynenko1977DamageBlistering}. In this model, a local pressurization is determined from the concentration of dissolved gas atoms, $c_g$, and the dissolution energy, $H$, of the gas species in the target material:
        \begin{equation}
            p = Hc_{g}
        \end{equation}
        A critical pressure for blistering onset is chosen as the pressure which exceeds the appropriate yield strength of the material, $\sigma_{y}$, resisting the overpressurization in implanted gas bubbles. A critical concentration, $c_c$, is the local gas concentration at which this pressure is exceeded. Local gas concentration is approximated by:
        \begin{equation}
            c_{g} = \frac{\Phi}{\Delta R}
        \end{equation}
        where $\Phi$ is the fluence and $\Delta R$ is the width of the distribution. The width of the distribution is determined from the width of the implantation distribution and the diffusion length:
        \begin{equation}
            \Delta R = \sqrt{\overline{\Delta R^{2}} + D t}
        \end{equation}
        Where $\overline{\Delta R^{2}}$ is the straggle or standard deviation of the implantation distribution, $D$ is the diffusion coefficient, and $t$ is the irradiation time. In the case of gas implantation via ion irradiation, it is often assumed that irradiation damage produces enough defects to serve as trapping sites that the diffusion coefficient is negligible \cite{Martynenko1977DamageBlistering} \cite{Scherzer1983DevelopmentImplantation}. Lattice defects caused by ion irradiation provide nucleation for fixed helium bubbles \cite{Kornelsen1972TheCrystal}. We can assume additional trapping due to the interaction of hydrogen with helium in the material, since helium bubbles serve as trapping sites for hydrogen atoms through synergistic effects \cite{Hayward2012SynergisticBubbles}, leading to a lower critical dose in the case of mixed hydrogen and helium exposure \cite{Guseva1981RadiationBlistering}. At temperatures well below the melting point, such as that expected for a relativistic interstellar spacecraft, high-flux hydrogen irradiation damage effects have been reproduced in low-flux experiments \cite{Gao2019High-fluxExposure}. However, for an interstellar probe, the flux is sufficiently low that diffusion effects may play a non-negligible role. Classical diffusion of individual gas atoms could lead to lower local gas atom concentrations as the implantation distributions widen, or increased gas atom concentrations around trapping sites such as grain boundaries. Helium bubble nucleation and migration could lead to increased gas atom concentrations and damage near surfaces, grain boundaries, and other defects \cite{Nakamura1977GrainBeams} \cite{Lane1983HeliumBoundaries} \cite{Goodhew1983OnBubbles}. No single theoretical framework exists to summarily treat these effects. However, simple models such as that presented here immediately offer compelling mitigation strategies. To perform this analysis, we find implantation profiles of ISM gas atoms at relativistic speeds using a BCA code, calculate critical concentrations for blistering onset for hydrogen and helium individually assuming a worst-case scenario of negligible diffusion, and show the effect of non-negligible diffusion on local gas concentrations.

\subsection{\label{sec:implantationdistributions}Ion Implantation Distributions}

        SRIM is a free-use but closed-source Monte Carlo, Binary Collision Approximation (BCA) code used to model ion-solid interactions in continuous development since 1985 \cite{Ziegler2010SRIM2010} \cite{Ziegler2004SRIM-2003}. A detailed description of the BCA model can be found in Eckstein \cite{Eckstein1991ComputerInteractions} and Robinson \cite{Robinson1994TheIntroduction}. For nuclear interactions, SRIM uses a universal interaction potential, the Ziegler-Biersack-Littmark potential, and the MAGIC algorithm to calculate the scattering angle of each binary collision \cite{Biersack1980ATargets}. To model electronic interactions, SRIM includes an electronic stopping formula that includes Lindhard-Scharff \cite{Lindhard1953EnergyCharge} electronic stopping at low energies (below approximately 25 keV/amu), Bethe-Bloch \cite{Ziegler1999StoppingMatter} stopping with corrections at high energies (above approximately 1 Mev/amu), and Andersen-Ziegler \cite{Andersen1977ReviewTheory} stopping at intermediate energies between the ranges of validity of Lindhard-Scharff and Bethe-Bloch. Electronic stopping is evaluated along each path between binary collisions with target atoms. SRIM produces detailed information about stopped ions, from which we find implantation distributions.

        \begin{figure}
            \centering
            \begin{tabular}{c}
                \includegraphics[width=0.5\textwidth]{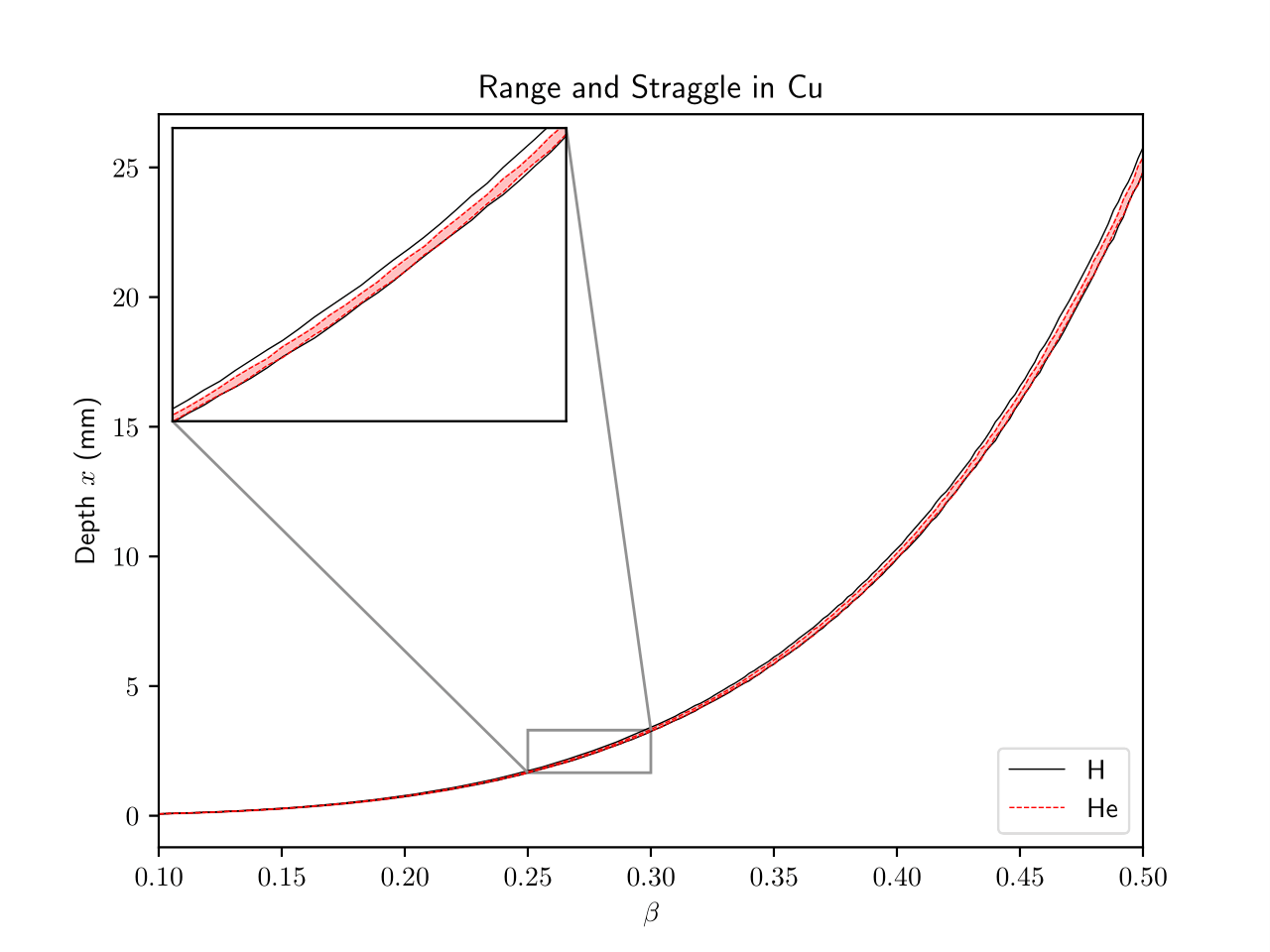} \\
                (a) \\
                \includegraphics[width=0.5\textwidth]{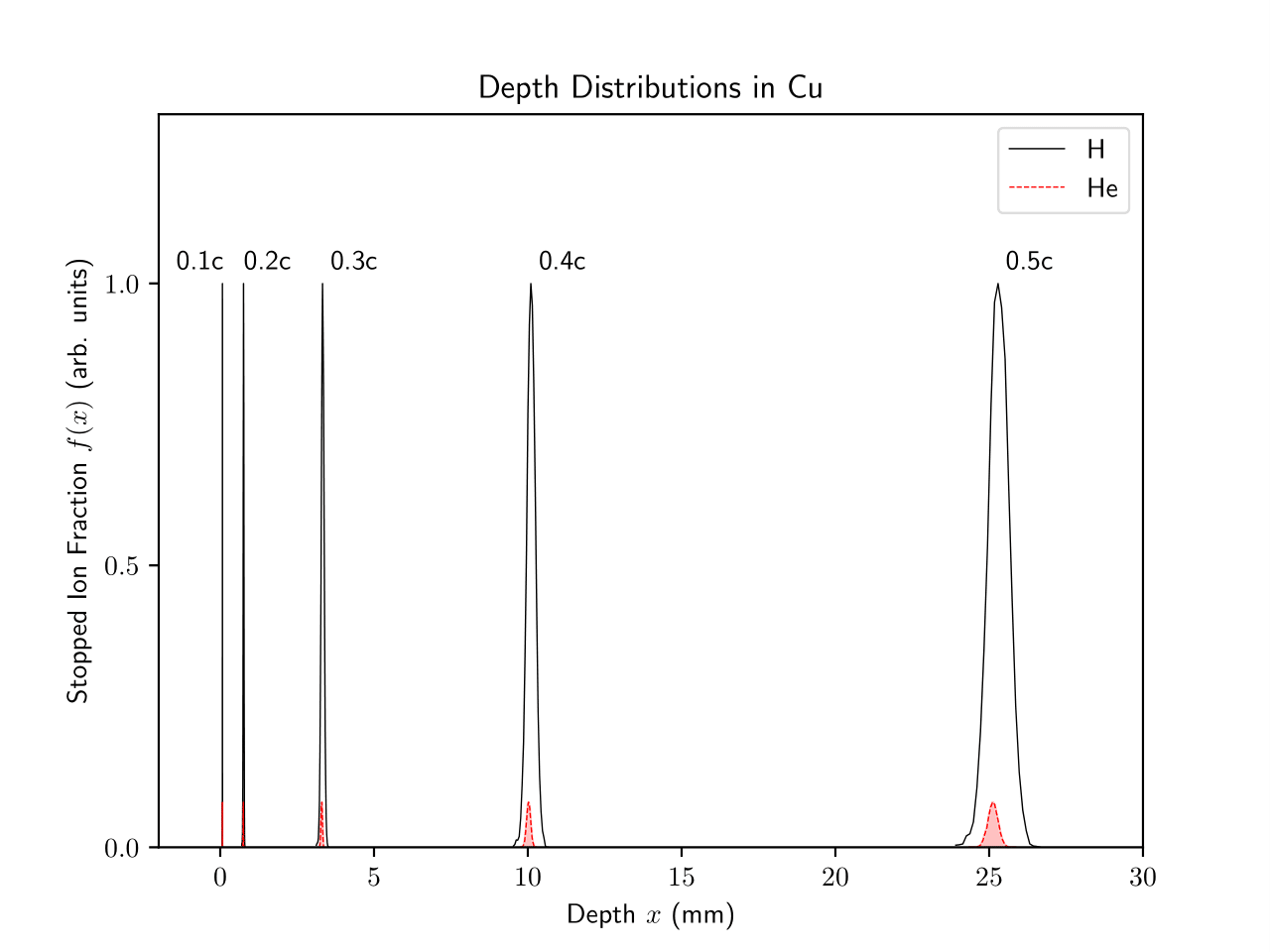} \\
                (b) \\
            \end{tabular}
            \caption{ \protect\centering (a) Implantation depths for H and He in copper from SRIM simulations. The regions bounded by solid lines for hydrogen and dashed lines for helium are centered on the mean implantation depth and span above and below the mean by the straggle, $\pm \Delta R$. (b) Implantation distributions, normalized and scaled to the appropriate ISM species number density, for H and He in copper from SRIM simulations at 0.1, 0.2, 0.3, 0.4, and 0.5c. Hydrogen distributions are unfilled and bounded by solid lines, and helium distributions are filled and bounded by dashed lines. Heights are normalized to 1 for all distributions.}
            \label{fig:srim_distributions}
        \end{figure}

        Although SRIM has recently come under scrutiny for inaccuracies for low to intermediate energy sputtering \cite{Shulga2019ASRIM} \cite{Shulga2018NoteSputtering} \cite{Wittmaack2004ReliabilityIons} and sputtered atom angular distributions \cite{Hofsass2014SimulationSRIM}, these issues do not affect the implantation distributions at high energy used in our analysis. From these ion implantation distributions we calculate the straggle, or standard deviation of the implantation distribution, and use this to estimate the local gas concentration in materials. For this calculation, ISM particles are assumed to be mono-energetic for the duration of the mission, and the acceleration phase is not considered.

        \begin{table}
        \caption{Range and straggle of hydrogen and helium at 0.2c}
        \label{tab:range}
        \begin{center}
            \begin{tabular}{c|c|c|c}
                 Material &  C (graphite) & Cu & Si\\
                 \hline
                 \hline                 $R_{\scriptsize{\textrm{H}}}$(0.2c) & 1.9mm & 0.72mm & 2.1mm\\
                 \hline
                 $R_{\scriptsize{\textrm{He}}}$(0.2c) & 1.9mm & 0.71mm & 2.1mm\\
                 \hline
                 $\overline{\Delta R^{2}}_{\scriptsize{\textrm{H}}}$(0.2c) & 30$\mu$m & 15$\mu$m & 35$\mu$m \\
                 \hline
                 $\overline{\Delta R^{2}}_{\scriptsize{\textrm{He}}}$(0.2c) & 13$\mu$m & 7.2$\mu$m & 14$\mu$m \\
            \end{tabular}
        \end{center}
        \end{table}

        Figure \ref{fig:srim_distributions} (a) shows the implantation depth and straggle of hydrogen and helium in copper, an example shield material, from 0.1 to 0.5c. Copper was chosen as a representative metallic shield material in which hydrogen is relatively soluble, potentially mitigating hydrogen blister formation \cite{Magnusson2017DiffusionCopper}. For velocities in excess of approximately 0.07c, the implantation distribution of helium out to one standard deviation from the mean fits entirely within one standard deviation of the hydrogen distribution. Inset is a zoomed in figure showing this overlap at velocities relevant to a relativistic, interstellar mission (0.25-0.3c). Any long duration, high velocity mission will be subject to a layer of mixed hydrogen and helium implanted many atomic layers deep below the surface. Figure \ref{fig:srim_distributions} (b) shows the full implantation distributions of hydrogen and helium in copper at 0.1, 0.2, 0.3, 0.4, and 0.5c from SRIM simulations, normalized to equal height. These results show that at speeds relevant to an interstellar mission, we expect a layer of mixed hydrogen and helium implanted in the spacecraft. Table \ref{tab:range} shows the results of SRIM simulations for a number of other example materials at 0.2c. In each material, hydrogen and helium stop at the same distance. As the velocity increases, the straggle increases, due to the increased number of small-angle atomic collisions incident particles are subject to as they travel deeper in the material.

\subsection{\label{sec:blisteringonset}Determination of Blistering Onset}

        A relativistic spacecraft on a mission to $\alpha$ Cen will likely be travelling between 10 and 30\% the speed of light. In the frame of the spacecraft, hydrogen from the interstellar medium will be impacting with an energy of 4.7-45.3 MeV, and helium will be impacting with an energy of 18.8-180.0 MeV. Assuming an average ISM number density of 1.0 cm$^{-3}$, with the hydrogen contributing 92\% and the helium contributing 8\% by number, the resulting flux for each species at 0.2c will be $2.8\times10^{13}$ s$^{-1}$m$^{-2}$ and $2.4\times10^{12}$ s$^{-1}$m$^{-2}$ for hydrogen and helium respectively. Over the course of a journey of 4.37 lyr, the approximate distance to $\alpha$ Cen, the fluence for hydrogen and helium will be $3.8\times10^{22}$ m$^{-2}$ and $3.3\times10^{21}$ m$^{-2}$ respectively, regardless of travel speed.

\begin{figure}
            \centering
            \includegraphics[width=0.47\textwidth]{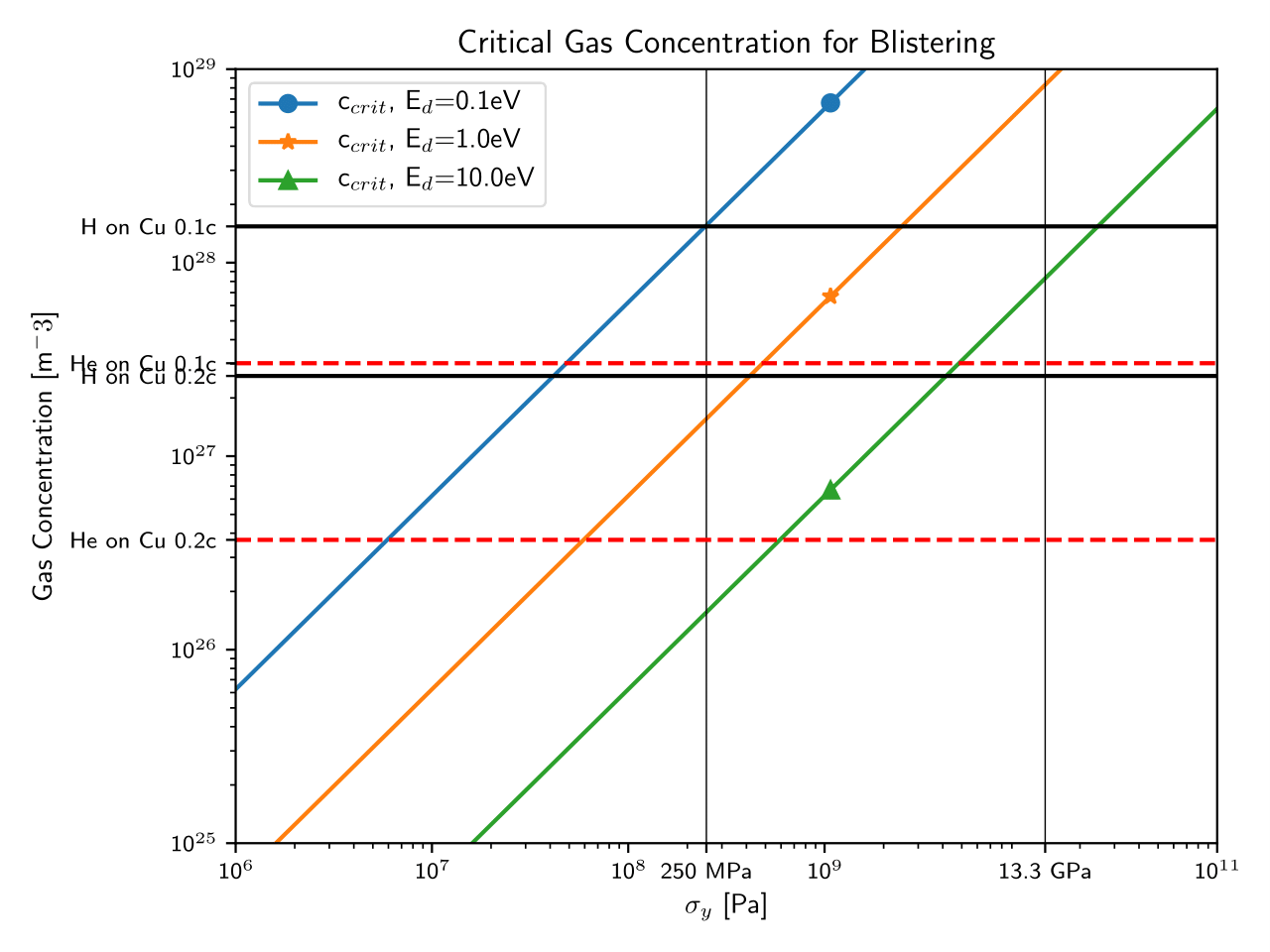}
            \caption{ \protect\centering Critical implanted gas concentrations for blistering onset calculated for hydrogen and helium assuming ISM particle density on the order of 1.0 cm$^{-3}$ for dissolution energies of 0.1, 1.0, and 10.0 eV in a material with yield strength $\sigma_{y}$. Horizontal lines marking estimated gas concentration in copper for hydrogen and helium at 0.1 and 0.2c are shown for reference. Vertical lines marking estimations of the yield strength against blistering in copper are shown. 250 MPa is the normal yield strength of copper. 13.3 GPa is 10\% of the elastic modulus of copper, per  \cite{Martynenko1977DamageBlistering}.}
            \label{fig:critical_dose}
        \end{figure}

        Using the model for determining blistering onset presented above, Figure \ref{fig:critical_dose} shows calculated critical concentrations of implanted gas atoms for a range of dissolution energies. Hydrogen, which dissolves readily in copper, has a dissolution energy near 0.5eV \cite{Magnusson2017DiffusionCopper}. For gas-material combinations with lower solubility, the dissolution energy will be significantly higher. Dissolution energies of hydrogen and helium are not widely available for many materials. Determining dissolution energies requires carefully designed experiments or computational modeling. Additionally, this calculation assumes a constant ISM number density of 1.0 cm$^{-3}$; if the actual number density is larger, the fluence and thus implanted gas concentrations will increase linearly with increasing number density. From Figure \ref{fig:critical_dose}, it is apparent that the local gas concentration of hydrogen and helium at 0.1c will be lower than at 0.2c. At higher relativistic velocities, incident ions will have travelled through significantly more material and will have experienced many more small-angle nuclear collisions, leading to increased straggle.

        At an ISM density of 1.0 cm$^{-3}$, neglecting diffusion and synergistic hydrogen-helium effects, and using the straggle of hydrogen and helium in copper, this model suggests that materials with a yield strength well in excess of 10 GPa will not blister. Determining the appropriate yield strength presents some difficulty. Using standard yield strength significantly underestimates the critical fluence for many materials; one strategy is to estimate the yield strength as 10\% of the elastic modulus \cite{Martynenko1977DamageBlistering}. Values for these quantities are reported for copper in Figure \ref{fig:critical_dose} as vertical lines at 250 MPa and 13.3 GPa, representing the traditional yield strength and 10\% of the elastic modulus respectively. Since we expect near-surface bubbles on surfaces perpendicular to ion exposure, choosing the correct yield strength will be critical for choosing target materials. This value may be obtainable from molecular dynamics simulations of near-surface bubbles. Note that this figure shows the threshold for blistering, and that bubble formation, swelling, and changes to material properties are expected at fluences many orders of magnitude lower than the blistering threshold, and these effects of gas accumulation will need to be mitigated.

\subsection{\label{sec:roleofdiffusion}Role of Diffusion in Cumulative Damage}
\begin{figure}
            \centering
            \includegraphics[width=0.5\textwidth]{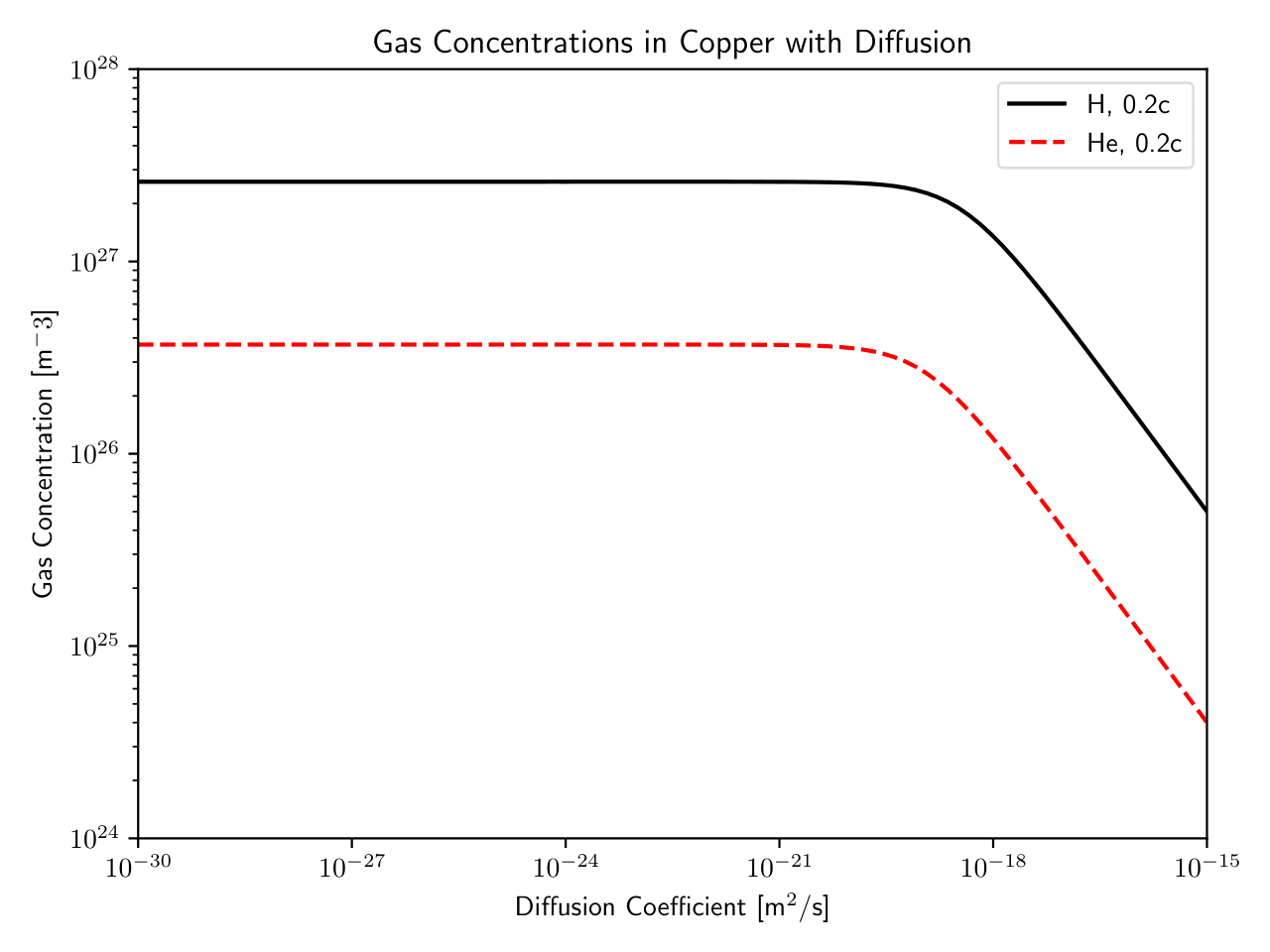}
            \caption{\protect\centering Gas concentration in copper after a journey to $\alpha$ Cen at 0.2c. Diffusion is only significant when the diffusion length becomes comparable to the width of the implantation distribution; in copper, this occurs when the diffusion coefficient is near 10$^{-17}$ m$^2$s$^{-1}$. This is on the order of classical diffusion coefficients in metals.}
            \label{fig:diffusion_coefficients_1}
        \end{figure}

        When our analysis extends to the case of significant classical gas species diffusion, we can determine for what diffusion coefficients diffusion will have a measurable effect on the local gas concentration. Figure \ref{fig:diffusion_coefficients_1} shows the estimated local gas concentration in a copper shield for a spacecraft traveling at 0.2c. Diffusion is only significant when the diffusion length becomes comparable to the straggle of the ion distributions. In copper, classical diffusion does not become significant until the diffusion coefficient exceeds 10$^{-17}$ m$^2$s$^{-1}$.

        This analysis assumes an isotropic diffusion coefficient for gas species, but implanted gas atoms can have significantly different transport behavior. For example, helium atoms implanted in many materials aggregate and form bubbles. Helium and hydrogen bubbles are known to have a synergistic effect in driving deleterious effects in  solids \cite{Hayward2012SynergisticBubbles}. This interaction can lead to increased local gas concentration, amplifying the effect of bubbles on the material.

        In a radiation damaged material, insoluble atoms may interact strongly with material defects such as vacancies, significantly limiting diffusion. For example, in body-centered cubic (BCC) metals such as tungsten, a candidate material for fusion reactor PFCs, the hydrogen diffusion activation energy is typically on the order of 0.1 eV, while the binding energy of a single vacancy site, capable of trapping as many as 10 atoms of hydrogen simultaneously, has been calculated to be as high as 1.2 eV \cite{Johnson2010HydrogenDecohesion}. This is a significant barrier to diffusion. Understanding the behavior of implanted gas atoms in spacecraft material including the effects of hydrogen-helium interactions, interactions with material defects, defect production by irradiation, and interaction of near-surface gas atoms with the material surface will be necessary to design materials resistant to bubble formation, cracking, and blistering.

        Molecular dynamics and density functional theory simulations are appropriate tools to study these effects in silicon, but in order to include realistic fluences at high energy in atomistic simulations, one must have access to significant computing resources, and artificially enhance the flux by many orders of magnitude to reach appropriate fluences in computable model times. Continuum modeling of bubble formation and bubble dynamics \cite{Blondel2018Continuum-scaleSurfaces} offers another approach, but relies on data from extensive HPC-scale MD calculations. Barring terrestrial experimentation, a simple analysis such as presented in this work, however, can be used to select promising shield materials with a significant safety margin to compensate for the lack of a complete multiscale model.

\section{\label{sec:discussion}Discussion}

\subsection{\label{sec:implications}Implications on Mission Success}

        The implications of the previously described phenomena are mission-threatening. For a 20-year journey without an adequate shielding system, hydrogen bubble formation and subsequent bursting may induce damage to the spacecraft's leading edge which, as the damage continues to make its way into the spacecraft body, will eventually damage components and electronics essential to the spacecraft's performance. Other spacecraft geometries, such as the needle-like spacecraft proposed by Lubin \cite{Lubin2016AFlight}, may help mitigate this issue by minimizing the front-facing surface area. If blistering and exfoliation does not occur, migration of gas atoms and bubbles into spacecraft components may lead to hardware failures. Additionally, over-pressurized bubbles bursting at the surface will both drastically alter the surface morphology and induce torques on the spacecraft, both of which will likely alter its long-term trajectory and orientation. Torques induced by bubble rupture must be compensated by the attitude control system (ACS), which might pose significant constraints on the design of the ACS and its authority over such perturbations. Moreover, secondary particle production and heavy species impacts have the potential to induce damage at a greater rate, depth, and severity than the common proton impact.

\subsection{\label{sec:mitigation}Mitigation Strategies}

        Various shielding schemes are proposed as mitigation strategies. Since the spacecraft is designed to fly edge-on into the ISM, a circumferential shield is proposed to protect the ISM-facing edge. The shield must satisfy three essential requisites: 1) block incoming proton radiation without structural failure due to hydrogen bubble formation, 2) stop any incoming heavier species, and 3) mitigate the production of secondary particles and their propagation through the spacecraft. Since the majority of the expected particle impacts will result in hydrogen build up at a localized depth below the surface, the Bragg peak, an effective shield would be at least two times as thick as the Bragg peak is deep.

        A favored shielding scheme consists of a sintered material at the H/He Bragg peak. The sintered nature of the material makes it granular with a characteristic grain and hole size that enable a high resistance to bubble formation \cite{Rossing1977ReductionReactors}. In a similar scheme, a porous ceramic material is used to surround the Bragg peak. The porous nature of the material acts similarly to the sintered material in that it makes it resistant to bubble formation and allows storage of the implanted hydrogen in its pores. A favorable material for the shield material is carbon due to its higher energy onset of secondary neutron production than most materials.

        Schemes involving metal powder layers surrounding the Bragg peak are predicted to perform in a similar fashion to the sintered material shields due to the granular nature of the material. Other shielding schemes involve layers of self-annealing metals that can be induced from solid to liquid states and back again repeatedly, such as gallium or mercury. In these schemes, incoming protons are collected at the Bragg peak while the metal is in a solid state. Then, once the rotation of the spacecraft carries a particular circumferential region out of the direction of the incoming radiation, the metal is heated until it liquefies, allowing the implanted hydrogen, in the form of gas atoms or hydrides, to diffuse to the edges and escape the spacecraft. Similar schemes involve the use of long-chained polymers, plastics, or circulating oils or liquids to mitigate bubble formation and store implanted hydrogen.

        \begin{figure}
            \centering
            \includegraphics[width=0.5\textwidth]{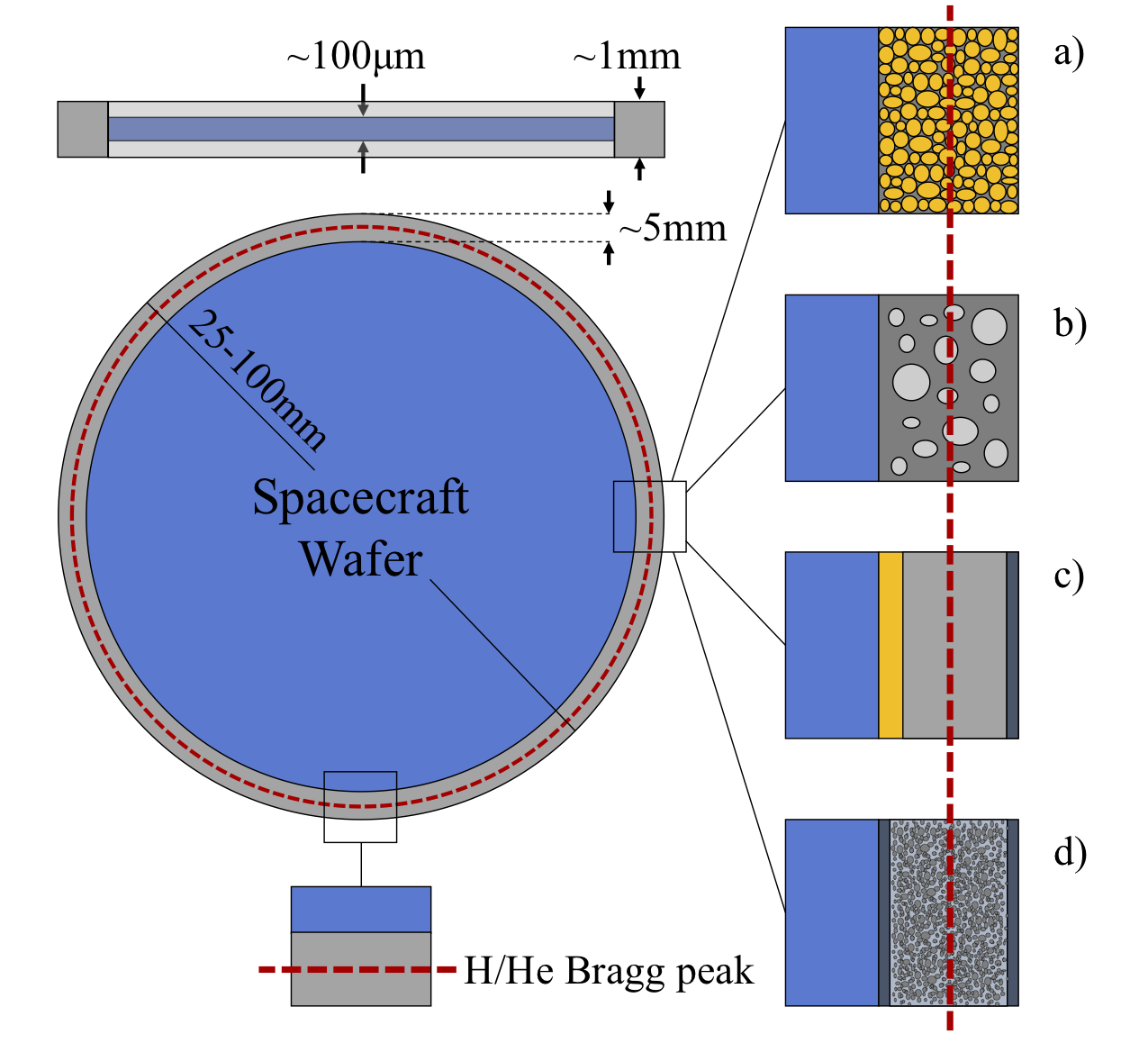}
           \caption{ \protect\centering Schematic of a wafer-like spacecraft geometry and four proposed shielding schemes. a) Sintered granular material, b) porous ceramic, c) self-annealing liquid layer with heat exchanger (yellow), d) metallic powder. Other shielding schemes involve plastics, oils, and/or circulating liquids.}
            \label{fig:shielding_schemes}
        \end{figure}

        More outlandish shielding schemes involve the deflection of incoming charged particles via electromagnetic fields about the circumference of the spacecraft. However, inherent issues such as high power consumption and strong field amplitudes limit the viability of this approach on a small spacecraft.

\subsection{\label{sec:experimentaldesign}Terrestrial Experimental Design}

        Terrestrial experimental investigation of the effect of ISM impacts on relativistic spacecraft will rely on an appropriate accelerator facility. Such a facility would need to meet the following requirements to be informative for this study:

        \begin{enumerate}
            \item High flux such that $\sim$20 years worth of ISM fluence can be simulated in hours, days, or weeks. Since we expect a total fluence of approximately 4$\times$10$^{22}$ m$^{-2}$, we would require a flux on the order of 10$^{16}$ m$^{-2}$s$^{-1}$ to achieve the target fluence with an exposure time of roughly 1.5 months.

            \item Ability to defocus the beam such that the ion flux can span the forward-facing cross-section of the target, investigating the possibility of blistering occurring on surfaces perpendicular to the direction of exposure.

            \item Cooling capabilities to control target temperature. At laboratory-relevant fluxes, we expect significantly higher target temperatures than for a relativistic spacecraft.

            \item Ability to implant hydrogen and helium at constant velocity. Simultaneous exposure may not be possible, so careful alternating exposures may be required.
        \end{enumerate}

        There do exist facilities that meet many of these requirements. Material diagnostics would include TEM/SEM/AFM of front-facing and perpendicular surfaces to determine surface morphology and characterize any swelling, blistering, or exfoliation, and FIB to measure depth-dependent composition to determine diffusion coefficients.

        \begin{figure}
            \centering
            \includegraphics[width=0.45\textwidth]{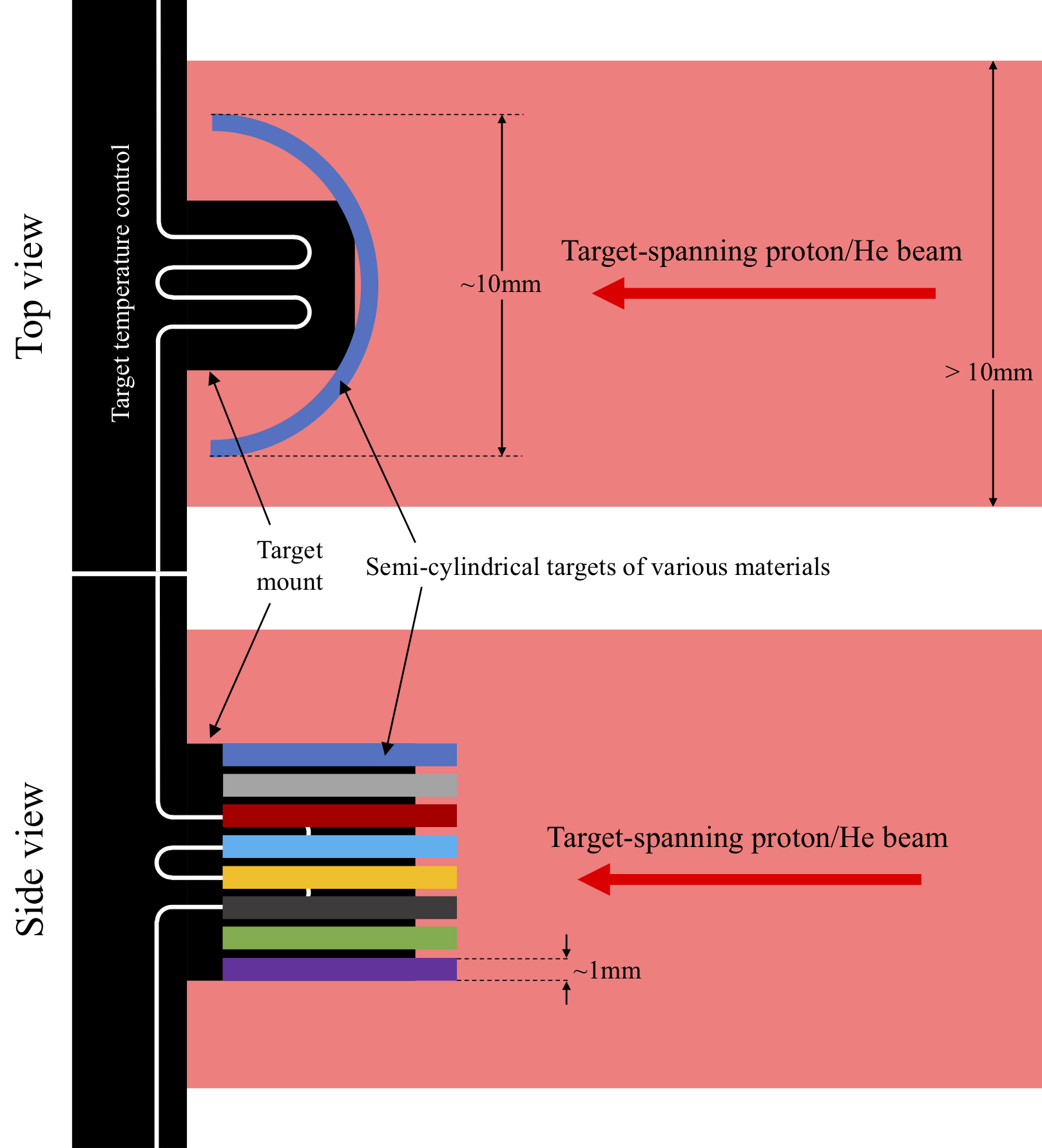}
            \caption{\centering Proposed experimental setup for radiation studies using a linear accelerator or cyclotron. Top: The target is to be fabricated in a cylindrical shape so as to study implantation over a continuous range of incident angles. Bottom: In the ideal scheme, the target mount would house multiple targets of various materials and the accelerator facility would have the capability to control the temperature of the target.}
            \label{fig:accelerator}
        \end{figure}

        Nuclear activation of the target due to proton irradiation is an additional concern for the spacecraft, but is also a concern for access to samples soon after experimental irradiations. A future paper will quantitatively address this issue.

\section{\label{sec:conclusion}Conclusion}

        In this work, we present the potentially mission critical phenomenon of damage to relativistic spacecraft by gas accumulation. Composed primarily of hydrogen and helium, exposure to the ISM at relativistic velocities will result in gas atoms accumulating many atomic layers below the front-facing surfaces of the spacecraft. Accumulation of hydrogen and helium in solids is associated with processes such as bubble formation, swelling, blistering, and exfoliation. Since the exposure will span the entire forward edge of the spacecraft, blistering may occur on surfaces perpendicular to the direction of travel. Using a zero-dimensional model of blistering onset, we have performed preliminary calculations in general and for copper, an example metallic shield material. For shield materials with a yield strength well in excess of 10 GPa, assuming an ISM number density of 1.0 cm$^{-3}$, this model predicts gas concentrations will remain below the critical concentration for blistering. Using an estimate of the appropriate yield strength against blistering as 10\% of the elastic modulus \cite{Martynenko1977DamageBlistering} and ignoring synergistic effects of mixed hydrogen and helium \cite{Hayward2012SynergisticBubbles}, this zero-dimensional model suggests that copper, an example metallic shield material, will blister from hydrogen implantation only if the ISM number density exceeds 60 cm$^{-3}$, significantly higher than any estimate, primarily due to the relatively high solubility of hydrogen in copper. However, neglected effects such as hydrogen-helium mixing and the proximity of stopped gas ions to surfaces perpendicular to the direction of exposure may lower the critical concentration enough for blistering to occur beyond the capabilities of the zero-dimensional model to determine.\\
        \indent Regardless of whether blistering occurs, the relatively high fluence suggests bubbles will form in the spacecraft shield. While no single theoretical framework exists to quantitatively explore bubble formation onset and the effects thereof, bubble formation in fusion PFCs provides a qualitative understanding of this phenomenon. Bubble formation leads to swelling and a decrease in the structural integrity of the spacecraft shield. Experiments or atomistic simulations of gas implantation and bubble formation with a wide-beam geometry need to be performed to make informed engineering decisions on shield materials for relativistic spacecraft.\\
        \indent Our zero-dimensional analysis suggests a number of strategies to mitigate bubble formation and blistering. First, increasing the diffusion coefficient, either through heating or material choice, will decrease the local gas concentration, preventing bubble formation and blistering. Second, a material that can accommodate large amounts of trapped gas, such as granular and sintered materials, will be resistant to blistering and swelling. Third, amorphous materials or liquid metals are particularly resistant to the effects of gas accumulation. Fourth, geometric optimization of the spacecraft such as high aspect ratio ``needle-like'' designs minimize the effective exposure area. Without accounting for the effects of bubble formation and potential blistering, spacecraft shields may erode or fail significantly sooner than instantaneous damage calculations would suggest.

        Total particle fluence depends only on the travel distance and ISM number density. There exists no feasible mechanism, for low mass spacecraft, by which ISM particles could be deflected at relativistic speeds, so any interstellar mission must have a mitigation strategy to deal with gas accumulation. Determining the ISM number density along the path of travel for an interstellar mission will be crucial for determining the fluence to which a relativistic spacecraft will be exposed. Blistering and exfoliation will erode the surface of any future missions to destinations significantly further than $\alpha$ Centauri without appropriate mitigating strategies. Regions of particularly high ISM number density will prove especially challenging to shield materials. Supercritical fluences for solids will require advanced mitigation strategies, such as self-healing, sintered, granular, or liquid metal shields that can accommodate the relentless accumulation of ISM particles incident upon future interstellar spacecraft traveling beyond $\alpha$ Centauri.

    \section{Diffusion}

    Effective diffusion coefficients in materials with a significant number of trapping sites can be estimated using an equilibrium trapping model \cite{Oriani1970TheSteel}. For a single kind of trapping site, the effective diffusion coefficient $D_{\text{eff}}$ is determined from the lattice diffusion coefficient, $D_L = D_0 \exp(-E_a/kT)$, the number of trapping sites per lattice site, $c_{T}/c_{L}$, the binding energy of the trapping site $E_b$, and the temperature $T$:

    \begin{equation}
        \label{eq:eff_diff}
        D_{\text{eff}} = \frac{D_0 \exp\left(-E_a/kT\right)}{1 + (c_T/c_L)\exp\left(E_b/kT\right)}
    \end{equation}

    To demonstrate the effect of radiation-induced vacancies on the effective diffusion coefficient for implanted gas atoms in the shield of an interstellar spacecraft, an example calculation for hydrogen incident upon a tungsten shield of a spacecraft travelling at 0.2c is shown. Tungsten was chosen because it has a relatively high hydrogen mobility, and 0.2c was chosen because it has the lowest vacancy density production per fluence rate, $R=4\times 10^{-5}$ vacancies per Angstrom-ion, of the three speeds considered in this work. The binding energy, $E_b$, for hydrogen trapping in tungsten vacancies is 1.4 eV \cite{Xu2009First-principlesCrystal}. $E_a$ and $D_0$ for hydrogen diffusion in pure tungsten are 0.21 eV and $5\times 10^{-8}$ m$^2$/s respectively \cite{Frauenfelder1969SolutionTungsten}. The straggle of hydrogen in tungsten at 0.2c is 20.72 $\mu$m, from Table  \ref{tab:straggle_02c}. Rearranging Equation \ref{eq:eff_diff} into Equation \ref{eq:fluence_limiting_diffusion}, the fluence for which enough vacancies are produced by the incident hydrogen at the effective diffusion length, $L_{\text{eff}} = \sqrt{D_{\text{eff}}t}$, is negligible (that is, below 1 $\mu$m) at 300 K  is $\Phi = 3.2\times 10^{3}$ cm$^{-2}$, a condition reached almost instantaneously when traveling at relativistic speeds through the ISM. Based on the negligible diffusion due to the effect of radiation-induced vacancies alone, diffusion will be considered negligible for the energies relevant to this study.

    \begin{equation}
        \label{eq:fluence_limiting_diffusion}
        \Phi = \frac{\left(D_0/L_{\text{eff}}^2\right) \exp \left(-E_a/kT\right) t - 1}{R \exp\left(E_b/kT\right)}
    \end{equation}

\section{Acknowledgements}
Funding for this program comes from NASA grants NIAC Phase I DEEP-IN – 2015 NNX15AL91G and NASA NIAC Phase II DEIS – 2016 NNX16AL32G and the NASA California Space Grant NASA NNX10AT93H and a generous gift from the Emmett and Gladys W. fund. We also acknowledge funding from ESA Contract No. 4000122836/18/NL/PS/gp and LLNL under Contract DE-AC52-07NA27344.
PML acknowledges support from the Breakthrough Foundation as a part of the Starshot program.

\bibliography{references}
\bibliographystyle{abbrv}



\end{document}